\documentclass{article}

\usepackage{arxiv}
\usepackage{changepage}

\usepackage[utf8]{inputenc} % allow utf-8 input
\usepackage[T1]{fontenc}    % use 8-bit T1 fonts
\usepackage{hyperref}       % hyperlinks
\usepackage{url}            % simple URL typesetting
\usepackage{booktabs}       % professional-quality tables
\usepackage{amsfonts}       % blackboard math symbols
\usepackage{nicefrac}       % compact symbols for 1/2, etc.
\usepackage{microtype}      % microtypography
\usepackage{lipsum}
\usepackage{graphicx}
\graphicspath{ {./images/} }

\title{QKD Entity Source Authentication: Defense-in-Depth for Post Quantum Cryptography}

\author{
 John J. Prisco \\
  CEO, Safe Quantum, Inc.,\\ Chairman of the QED-C Use Cases Technical Advisory Committee \\
  \texttt{john.prisco@safequantum.com} \\
  }

\begin{document}
\maketitle
\begin{abstract}
Quantum key distribution (QKD) was conceived by Charles Bennett and Gilles Brassard in December of 1984.\cite{Bennett1984}  In the ensuing 39 years QKD systems have been deployed around the world to provide secure encryption for terrestrial as well as satellite communication.  In 2016 the National Institute of Standards and Technology (NIST) began a program to standardize a series of quantum resistant algorithms to replace our current encryption standards thereby protecting against future quantum computers breaking public key cryptography.  This program is known as post quantum cryptography or PQC. One of the tenets of cybersecurity is to use an approach that simultaneously provides multiple protections known as defense-in-depth.  This approach seeks to avoid single points of failure.  The goal of this paper is to examine the suitability of a hybrid QKD / PQC defense-in-depth strategy. A focus of the paper will be to examine the sufficiency of initial QKD hardware authentication (entity source authentication) which is necessary to guard against man-in-the-middle attacks. 
\end{abstract}

% keywords can be removed

%%%%%%%%%%%%%%%%%%%%%%%%%%  body  %%%%%%%%%%%%%%%%%%%%%%%%%%
\section{Executive Summary}

\begin{itemize}
\item Today the security of our information systems is more important than ever for personal, economic, and national security reasons. At the same time, attackers, both individual and state sponsored, are continually seeking and finding 
vulnerabilities.

\item Encryption techniques based on factoring are widely used to protect online transactions. However, Shor’s algorithm for factoring large numbers can break this type of encryption once a quantum computer that is sufficiently powerful, also known as a cryptographically relevant quantum computer (CRQC), is available.

\item One approach to address the threat posed by CRQCs is to implement new cryptographic standards that are, to the best of our knowledge, not vulnerable.  NIST, in coordination with the National Security Agency (NSA), is focused on selecting PQC algorithms (which are quantum-resistant mathematical algorithms) to become the new standard for quantum-resistant encryption. 

\item This project has been conducted in a public vetting process since 2016 and is expected to conclude with a new standard within the next year. To select algorithms with more diversity than the current batch of lattice-based algorithms, NIST reopened its submission process for new proposals last year and recently reported that it had received 40 new qualified proposals before its June 1, 2023 submission deadline. 

\item Another approach to address the threat of CRQCs is to take advantage of characteristics of quantum mechanics that enable the detection of the presence of an eavesdropper and therefore provide assurance of secure exchange of information. This approach is known as Quantum Key Distribution (QKD). 

\item QKD is a secure communication method for exchanging encryption keys only known between shared parties. It uses properties found in quantum physics to exchange cryptographic keys in such a way that is provable and guarantees security. QKD enables two parties to produce and share a key that is used to encrypt and decrypt messages. Specifically, QKD is the method of distributing the keys between parties.\cite{TechTarget} Benefits of QKD include a reduced vulnerability to increasing computational power and an immediate detection of the presence of eavesdroppers.

\end{itemize}

\noindent Known limitations of QKD include those described by National Security Agency researchers who in 2020 published a paper entitled “Quantum Key Distribution (QKD) and Quantum Cryptography (QC)” \cite{NSA} in which they describe five reasons why QKD needs further study and development.

\begin{enumerate}
\item Quantum key distribution is only a partial solution lacking hardware authentication.
\item Quantum key distribution requires special purpose equipment.
\item Quantum key distribution increases infrastructure costs and insider threat risks.
\item Securing and validating quantum key distribution is a significant challenge.
\item Quantum key distribution increases the risk of denial-of-service attacks.
\end{enumerate}

\noindent In this paper we primarily examine and propose a defensible QKD authentication solution to address, in part, the concerns raised by NSA.  Our ultimate goal is to add QKD to a post quantum computer defense-in-depth strategy alongside PQC.  Achieving this goal will require time and development of QKD performance.  Currently, there are no QKD resources at NIST/NSA paralleling the PQC effort. We advocate for them to join in the advancement of QKD and to gain their endorsement. 

We recommend a paper \cite{renner2023} written by Renato Renner and Ramona Wolf in the Arxiv: entitled “The debate over QKD: A rebuttal to the NSA's objections” (arxiv.org). This paper offers thoughts about overcoming NSA’s concerns with QKD.

\medskip

\section{Importance of authentication for QKD security}
When implementing secure authentication for QKD systems, sufficient care must be taken to achieve information-theoretic security for the system. If the authentication key is not information-theoretically secure, an attacker can break it and thereby bring all classical and quantum communications under the attacker’s control with the objective of relaying the information by means of a man-in-the-middle attack.

In the NSA paper previously sited, there is concern with initial authentication or “entity source authentication”. For QKD to be secure, the classical messages used in forming the key must be authenticated.  It is not necessary to authenticate the qubit communication.
Quantum mechanics offers information security assurances that cannot be achieved classically, such as preserving the confidentiality and privacy of communications when it is intercepted by sophisticated adversaries with virtually unlimited computing resources. 
Due to the “no cloning” theorem of quantum mechanics, it is not possible to intercept quantum communications without the destruction of the information carried, leading to noticeable reduction of communication quality and detection by the communicating parties.  

With the technology currently available, each quantum transmitter and receiver must be specifically tuned to function properly.  If an injection attack or man-in-the-middle attack were to occur in the quantum channel, then the network operators would immediately become aware because the affected network links would be out of tolerance upon the change and most likely stop working all together.

However, there still exists potential attack scenarios where sophisticated attackers may be able to spoof, tamper or exploit QKD systems to compromise its communication.
In order to guard against potential attack scenarios, it is essential to securely provide initial authentication or “entity source authentication” of the QKD hardware. For QKD to be secure, the classical messages used in forming the keys must be authenticated.

\subsection{\textbf{Information theoretical authentication}}
The QKD protocol that we are examining is BB84 protocol developed by Charles Bennett and Gilles Brassard in December 1984. The following, in italics, is from their paper describing the initial entity source authentication which we rely on in our approach to identify Alice and Bob hardware for a QKD system. 

\medskip
\begin{adjustwidth}{3em}{3em}
\emph{In quantum public key distribution, the quantum channel is not used directly to send meaningful messages, but is rather used to transmit a supply of random bits between two users who share no secret information initially, in such a way that the users, by subsequent consultation over an ordinary non-quantum channel subject to passive eavesdropping, can tell with high probability whether the original quantum transmission has been disturbed in transit, as it would be by an eavesdropper (it is the quantum channel's peculiar virtue to compel eavesdropping to be active). If the transmission has not been disturbed, they agree to use these shared secret bits in the well-known way as a one-time pad to conceal the meaning of subsequent meaningful communications, or for other cryptographic applications (e.g., authentication tags) requiring shared secret random information. If transmission has been disturbed, they discard it and try again, deferring any meaningful communications until they have succeeded in transmitting enough random bits through the quantum channel to serve as a one-time pad.
The need for the public (non-quantum) channel in this scheme to be immune to active eavesdropping can be relaxed if the Alice and Bob have agreed beforehand on a small secret key, which they use to create Wegman-Carter authentication tags for their messages over the public channel. In more detail the Wegman-Carter multiple message authentication scheme uses a small random key to produce a message-dependent 'tag' (rather like a check sum) for an arbitrary large message, in such a way that an eavesdropper, ignorant of the key, has only a small probability of being able to generate any other valid message-tag pairs. The tag thus provides evidence that the message is legitimate, and was not generated or altered by someone ignorant of the key. (Key bits are gradually used up in the Wegman-Carter scheme, and cannot be reused without compromising the system's provable security; however, in the present application, these key bits can be replaced by fresh random bits successfully transmitted through the quantum channel.) The eavesdropper can still prevent communication by suppressing messages in the public channel, as of course he can by suppressing or excessively perturbing the photons sent through the quantum channel. However, in either case, Alice and Bob will conclude with high probability that their secret communications are being suppressed, and will not be fooled into thinking their communications are secure when in fact they're not.}
\end{adjustwidth}
\medskip

\noindent This is meant to solve the problem of “we have negotiated a key securely over the quantum channel, but we don’t know who we have negotiated a key with” because one now knows who specifically the bits came from through the Wegman-Carter authentication tags described in the above excerpt from the Bennett and Brassard paper.
This solution works best for point-to-point links or simple QKD networks, with a small number of participants. It offers the best possible security. However, when the number of potential pairs of users increases, it can quickly become too complex. This is why another approach can be used.

\subsection{\textbf{Authentication with hash-based signatures}}

The most secure authentication algorithms, which do not rely on the complex mathematical problems needed by other PQC algorithms are the hash-based signatures. The security of hash-based signature relies on the existence of one-way functions alone (and, of course, on correct implementations). Several candidates have been considered. Stateful signatures are probably the fastest and smallest, but, since the users must keep track of each signature, in order to prevent re-use of the same keys, these are plagued with serious implementation issues. Therefore stateless signatures, such as Sphincs+, have been proposed. Sphincs+ is one of the signatures algorithms under standardization from NIST. 

\subsection{\textbf{Authentication with other PQC signatures}}
The other signature schemes under consideration rely on the existence of a one-way function with trapdoors. The immediate question, which may be asked when relying on PQC signatures for authenticating QKD, is that, if we need PQC for signatures, why do we need QKD at all? We could indeed use PQC for both authentication and key exchange. The reason why we believe that PQC for signature and QKD for key exchange still makes complete sense is a question of timing. Authentication has to be safe up to the time of the key exchange. On the other hand, confidentiality has to be safe after the exchange, until the time of validity of the encrypted information. If we trust an authentication PQC algorithm to be safe today, we can safely use it. We have to be sure that a key exchange PQC algorithm stays safe for a potentially much longer period. This is what is offered by QKD.  Moreover, having defense-in-depth with QKD and PQC eliminates a single point of failure.

After reviewing our approaches with cryptography experts from MITRE, the University of Chicago and the University of South Florida we settled upon an authentication approach which uses the PQC standards finalist CRYSTALS Dilithium. We believe this to be an excellent choice because we are proposing a hybrid of PQC and QKD to solve the authentication concerns and achieve the long-term confidentiality offered by QKD.

\section{Conclusions}

No one can afford to rebuff any quantum-safe security solution. Europe, Japan and China have robust QKD systems and trials in place. Shouldn’t the United States be as informed as our international friends and foes?

No security expert would ever recommend to use just one system or rely on just one approach. A defense-in-depth strategy deploying more than one kind of encryption can provide the layers of protection that will prove resilient. 

QKD Entity Source Authenticated is an important part of quantum security. QKD Entity Source Authentication, as described in the proposed protocol of this paper, solves a major concern in the overall security of QKD.
Currently, there are no QKD resources at NIST/NSA paralleling the PQC effort. We advocate for them to join in the advancement of QKD and gain their endorsement.

\section{Acknowledgments}
I would like to thank the following individuals who contributed greatly to this paper. 

\noindent \textbf{Carl Dukatz}, Computing Subcommittee Lead of the QED-C Use Cases TAC and Accenture Managing Director-Global Lead for Quantum Computing at Accenture, \textbf{William Trost}, Member of the QED-C Communications and Security Subcommittee of the Use Cases TAC and Lead Member of Technical Staff at AT\&T, Chief Security Office (CSO), Quantum Computing Cybersecurity Initiative, \textbf{Kirk McGregor}, Member of the QED-C Communications and Security Subcommittee of the Use Cases TAC and Chief Strategy Officer at Iff Technologies, and affiliate researcher with the Lung Center in the Department of Internal Medicine, School of Medicine at the UC Davis Medical Center and with Expolab in the Department of Computer Science at UC Davis. \textbf{Elizabeth Wood}, Member of the QED-C Communications and Security Subcommittee of the Use Cases TAC and Professional in the AT\&T Chief Security Office (CSO), Quantum Computing Cybersecurity Initiative.
\medskip

\noindent Special thanks to \textbf{Celia Merzbacher}, Executive Director of the Quantum Economic Development Consortium (QED-C) for her helpful review of this paper.

\bibliographystyle{unsrt}  
\bibliography{references}  %%% Remove comment to use the external .bib file (using bibtex).

\end{document}